\documentclass{ws-ijmpd}

\usepackage{lscape}
\newcommand{\be}{\begin{equation}}
\newcommand{\ee}{\end{equation}}
\newcommand{\bea}{\begin{eqnarray}}
\newcommand{\eea}{\end{eqnarray}}

\begin{document}

\markboth{L\'opez-Corredoira et al.}
{The QSO Hubble Diagram}

%
\catchline{}{}{}{}{}
%

\title{Cosmological test with the QSO Hubble diagram}

\author{M. L\'opez-Corredoira$^{1,2}$, F. Melia$^3$, E. Lusso$^4$ and G. Risaliti$^{5,4}$}

\address{$^1$ Instituto de Astrof\'\i sica de Canarias,\\ 
E-38205 La Laguna, Tenerife, Spain\\
$^2$ Departamento de Astrof\'\i sica, Universidad de La Laguna,\\
E-38206 La Laguna, Tenerife, Spain;\\
$^3$ Department of Physics, The Applied Math Program, and Department of Astronomy,
The University of Arizona, AZ 85721, USA;\\
$^4$ INAF - Arcetri Astrophysical Observatory, Largo E. Fermi 5,
I-50125 Firenze, Italy;\\
$^5$ Dipartimento di Fisica e Astronomia, Universit\`a di Firenze, via G. Sansone 1, 50019 Sesto Fiorentino (Firenze), Italy}

\maketitle

\begin{history}
\received{Day Month Year}
\revised{Day Month Year}
\comby{Managing Editor}
\end{history}

\begin{abstract}
A Hubble diagram (HD) has recently been constructed in the redshift range
$0\lesssim z\lesssim 6.5$ using a non-linear relation between the ultraviolet and 
X-ray luminosities of QSOs. The Type Ia SN HD has already provided a high-precision 
test of cosmological models, but the fact that the QSO distribution extends well
beyond the supernova range ($z\lesssim 1.8$), in principle provides us with an 
important complementary diagnostic whose significantly greater leverage in $z$ 
can impose tighter constraints on the distance versus redshift relationship. In 
this paper, we therefore perform an independent test of nine different cosmological 
models, among which six are expanding, while three are static. Many of these are
disfavoured by other kinds of observations (including the aforementioned Type Ia
SNe). We wish to examine whether the QSO HD confirms or rejects these earlier 
conclusions. We find that four of these models (Einstein-de Sitter, the Milne 
universe, the Static Universe with simple tired light and the Static universe 
with plasma tired light) are excluded at the $>99\%$ C.L. The Quasi-Steady State
Model is excluded at $>95$\% C.L. The remaining four models ($\Lambda$CDM/$w$CDM, 
the $R_{\rm h}=ct$ Universe, the Friedmann open universe and a Static universe 
with a linear Hubble law) all pass the test. However, only $\Lambda$CDM/$w$CDM 
and $R_{\rm h}=ct$ also pass the Alcock-Paczy\'nski (AP) test. The optimized 
parameters in $\Lambda$CDM/$w$CDM are $\Omega _m=0.20^{+0.24}_{-0.20}$ and 
$w_{de}=-1.2^{+1.6}_{-\infty }$ (the dark-energy equation-of-state). Combined 
with the AP test, these values become $\Omega _m=0.38^{+0.20}_{-0.19}$ and 
$w_{de}=-0.28^{+0.52}_{-0.40}$. But whereas this optimization of parameters 
in $\Lambda$CDM/$w$CDM creates some tension with their concordance values, 
the $R_{\rm h}=ct$ Universe has the advantage of fitting the QSO and AP data 
without any free parameters.
\end{abstract}



\section{Introduction}
\label{.intro}

Cosmological models with a geometry different from that in the
current standard model have fallen out of favour and are rarely
considered in ongoing tests using the latest high-precision
measurements.  However, even within the framework of the
standard model, not all the data fit together tension free. At least
some controversy still surrounds the interpretation of various
measurements, and other competing models often fit at least some
of these observations better than the concordance model does.\cite{Lop14a}
It is therefore useful to re-examine how these alternative scenarios fare
compared to $\Lambda $CDM when new, improved data become
available. The principal reason is that if the latest observations
strongly confirm the reasons they were disfavoured in the first
place, this can only solidify the concordance model's status as
the correct model of the Universe. In addition, there is the
possibility that something may have been missed.

The cosmological measurements that shed light on the geometry of the
Universe may be separated into two principal categories. The first
includes a measurement of the fluctuations in the Cosmic Microwave Background
Radiation (CMB) and the analysis of large-scale structure via the inferred
distribution of galaxies. CMB anisotropies provide the most evident support
for the concordance ($\Lambda$CDM) model, but one should find an independent
confirmation of this theory and its parameters because CMB anisotropies
may be generated/modified by mechanisms other than those in the standard picture,\cite{Nar07,Ang11,Lop13,Mel14a} and may also have some contamination.\cite{Lop07}

Secondly, cosmological tests using surveys of galaxies have also been developed to
provide information on the geometry of the Universe. These include the angular-size 
test;\cite{Kap87,Lop10} the surface brightness (known as the
`Tolman' \cite{Lub01,Ler14}) test; the use of Hubble diagrams (HD)
for galaxies;\cite{Sch97}; and Gamma-Ray Bursts.\cite{Wei10,Wei13,Wei14a} However, these depend on the evolution of the sources, so the results of this 
type of test may vary hugely depending on one's interpretation. A better prospect is
obtained with the Alcock-Paczy\'nski test,\cite{Lop14b,Mel15c} which can test the
geometry of the Universe independently of the evolution of galaxies.

Another good prospect is based on the use of an HD constructed
from Type Ia Supernovae (SNIa) embedded within the galaxies,\cite{Kow08} provided that we assume zero evolution and negligible extinction or 
selection effects (which are not universally accepted; see, e.g., criticisms by Refs.
\cite{Bal06,Pod08,Bog11}). Even
then, these events are detectable only up to redshift $z\sim 2$.\cite{Jon13}

Recently, a new method was presented\cite{Ris15} 
of sampling the redshift-distance relationship, based on a non-linear
correlation between the ultraviolet and X-ray luminosities of Quasi Stellar
Objects (QSOs). This relationship appears to be independent of evolution, so
suitable sources may be found out to a redshift of six. This new diagnostic
was used by these authors to optimize the parameters in the standard model.
In this {\it letter}, we present a followup application, based on the same data
described in Ref. \cite{Ris15}, to carry out a comparative analysis of nine different
cosmological models, six of them with expansion and three representing
a static Universe.
Exotic models invoking a static Universe are included for the
simple reason that the discussion on the reality of the expansion is still 
being discussed within some literature.\cite{Lop15} The principal reason for
including them here is not to resurrect them but, rather, to make
use of this excellent new diagnostic tool to re-test them against
the standard model independently of previous studies. In principle, since 
this diagnostic is based on a distribution of sources extending over a much 
larger range in redshift than, say, the Type Ia SNe, it has the potential
of providing tighter constraints than are currently available. Such an 
independent test should be viewed as a complement to current
consensus, in the sense that if this test also disfavours models already 
disfavoured by other observations, then their case is further weakened
in comparison to models that are favoured.

\section{Competing cosmological scenarios}
\label{.cosmomodels}
We will test nine different cosmological models, each with its unique
expression for the luminosity distance, $d_L(z)$. We assume a Hubble constant
$H_0=70$ km s$^{-1}$ Mpc$^{-1}$ throughout.

\begin{enumerate}

\item The flat (concordance) $\Lambda$CDM model, characterized by
the parameters $\Omega_m=0.3$, $\Omega_\Lambda=0.7$ and a dark-energy
equation-of-state $w_\Lambda =-1$. Here, $\Omega_i$ is the energy density
$\rho_i$ of species $i$, scaled to today's critical density, $\rho_c\equiv
3c^2H_0^2/8\pi G$. In this model,
\begin{equation}
d_L(z)=\frac{c}{H_0}(1+z)
\int _0^z\frac{du}{\sqrt{\Omega _m(1+u)^3
+\Omega _\Lambda (1+u)^{3(1+w_\Lambda )}}}\;.\label{concordance}
\end{equation}

\item Einstein--de Sitter (essentially Eq.~1 with $\Omega_m=1$ and
$\Omega_\Lambda=0$):
\begin{equation}
d_L(z)=2\frac{c}{H_0}(1+z)\left(1-\frac{1}{\sqrt{1+z}}\right) .
\end{equation}
Although this is no longer the standard model, some researchers
still view it as more appropriate than the concordance model.\cite{Vau03,Bla06}
\vskip 0.1in

\item A Friedmann model with negative curvature. Here, $\Omega_m=0.3$,
$\Omega_\Lambda=0$, implying a curvature term with $\Omega_K=1-\Omega _m=0.7$
(Ref. \cite{Bar12}[\S 7.4.1]):
\begin{equation}
d_L(z)=\frac{c}{H_0}\frac{(1+z)}{\sqrt{\Omega _K}}
\sinh \left(\int _0^z\frac{\sqrt{\Omega_K}\,du}{\sqrt{\Omega _m(1+u)^3
+\Omega _K(1+u)^2 }}\right)
.\end{equation}
Such a model might be relevant when one wishes to avoid including a
cosmological constant (i.e., $\Omega_\Lambda=0$).
\vskip 0.1in

\item The Quasi-steady State Cosmology (QSSC \cite{Nar07}):
\begin{equation}
d_L(z)=\frac{c}{H_0}(1+z)
\int _0^z\frac{du}{\sqrt{\Omega _c(1+u)^4+\Omega _m(1+u)^3
+\Omega _\Lambda }}\;.
\end{equation}
This cosmology is not the standard model, but has been used to fit
an assortment of HDs, e.g., for SNIa.\cite{Ban00,Nar02,Vis02} The expansion with an
oscillatory term produces a dependence of the luminosity and angular-diameter
distances similar to those of the standard model, though adding the effects
of matter creation (the so-called C-field, for which $\Omega_c=1-\Omega_m-
\Omega_\Lambda$) with slight variations depending on the parameters,
which are not as well constrained as those in the standard model.
Here, we will keep $\Omega _m$ and $\Omega _\Lambda $ as unconstrained
parameters. Previous estimates of these variables imply that galaxies
should only be observable out to a maximum redshift of $z\lesssim 6$,
a result that is already incompatible with galaxies observed at redshift 8
and beyond.\cite{Leh10}
\vskip 0.1in
\item The $R_{\rm h}=ct$ Universe (a Friedmann-Robertson-Walker cosmology
with zero active mass). This model has a total equation-of-state
$\rho+3p=0$, where $\rho$ and $p$ are, respectively, the total energy
density and pressure of the cosmic fluid.\cite{Mel07,Mel12,Mel16}
In this cosmology,
\begin{equation}
d_L(z)=\frac{c}{H_0}(1+z)\ln (1+z) .
\end{equation}
\vskip 0.1in

\item The Milne Universe. This solution may be obtained from the more generic
FRW equations by demanding that the energy density, pressure and cosmological
constant are all equal to zero and the spatial curvature is negative ($k=-1$). 
From these assumptions, and the Friedmann equations, it follows that the scale
factor must depend linearly on time. In the model, the mathematical equivalence
of the zero energy density ($\rho = 0$) version of the FRW metric with Milne's
model implies that a full general relativistic treatment using Milne's assumptions
would result in an increasing scale factor and associated metric expansion of
space, with the feature of a linearly increasing scale factor for all time.\cite{Vis13,Cha15}. 
The luminosity distance is\cite{Vis13}
\begin{equation}
d_L(z)=\frac{c}{H_0}(1+z)\sinh\left[{\rm ln} (1+z)\right].
\end{equation}
\vskip 0.1in

\item A static Euclidean model with a linear Hubble law at all redshifts:
\begin{equation}
d_L(z)=\frac{c}{H_0}\sqrt{1+z}\,z\label{angdistst}.
\end{equation}
This model has been used by some authors to account for certain specific
observations.\cite{Ler14} This scenario assumes that the
Universe is static. The factor $\sqrt{1+z}$ stems from the loss of energy
due to a redshift without expansion. This factor is different from $(1+z)$
because there is no time dilation. The challenge with this model is to
account for the redshift using a mechanism different from the conventional
expansion/Doppler effect. This cosmology has not been explored theoretically
and/or mathematically, but its promoters argue that, from a phenomenological
point of view, one may consider this relationship between distance and redshift
as an extrapolation of the observed behavior at low redshifts. In this
paper, our goal is simply to test its predictions against the QSO data,
independently of how or why one may justify its theoretical basis.
\vskip 0.1in

\item A static Euclidean model with tired light:
\begin{equation}
d_L(z)=\frac{c}{H_0}\sqrt{1+z}\,{\rm ln} (1+z). 
\label{angdisttl}
\end{equation}
This phenomenological representation stems from the idea that photons
lose energy along their trajectory due to some interaction, and the
relative loss of energy is proportional to the path length
i.e., $\frac{dE}{dr}=-\frac{H_0}{c}E$.\cite{LaV12}(\S 7.3). 
Of course, as in the previous case, this ansatz does not
enjoy much support among cosmologists, but our goal here is merely
to test its predictions agains the QSO data.
\vskip 0.1in

\item A static Euclidean model with plasma tired light:
\begin{equation}
d_L(z)=\frac{c}{H_0}(1+z)^{3/2}\,\ln (1+z).
\end{equation}
The plasma redshift application\cite{Bry04}(\S 5.8)
introduces a factor $(1+z)^{3/2}$ instead of $(1+z)^{1/2}$ to take into
account an additional Compton scattering which is double that of the plasma
redshift absorption.
\end{enumerate}

\section{The QSO HD}

A nonlinear relation between the rest-frame ultraviolet (2500 \AA )
and X-ray  (2 keV) luminosities of quasars,
of the type $\log L_X=\beta +\gamma \log L_{\rm UV}$ (e.g. Refs. \cite{steffen06,just07,lusso10}), is what
allows to derive a Hubble diagram for these sources.
Equation~(5) in Ref. \cite{Ris15} relates the rest-frame ultraviolet and X-ray fluxes of QSOs
according to
\begin{equation}
DM(z)=\frac{5}{2(\gamma -1)}[\log _{10}(F_{\rm X})-\gamma \log _{10}(F_{\rm UV})-\beta ']\,,
\end{equation}
where $\gamma $  and $\beta '(=\beta+(\gamma -1)\log_ {10}(4\pi )$ are constants (in principle independent of $z$ if there
is no evolution), and
\begin{equation}
DM(z)\equiv 5\log _{10}[d_L(z/{\rm Mpc})]-25
\end{equation}
is the distance modulus. In their analysis, Ref. \cite{Ris15}(\S 4) also determined that
$\gamma =0.60\pm 0.02$ for all $z$, and this result is independent of the cosmological
model since it only depends on the fluxes at rest that are derived from the observations. Hence,
\begin{equation}
DM(z)+A=-(6.25\pm 0.31)
[\log _{10}(F_{\rm X})-(0.60\pm 0.02) \log _{10}(F_{\rm UV})]\,,
\end{equation}
where $A$ is an arbitrary scaling factor. This is the relationship
that we will use to examine how well the expressions for $d_L$ predicted
by the various cosmologies introduced in \S~2 fit the data.

\begin{figure}
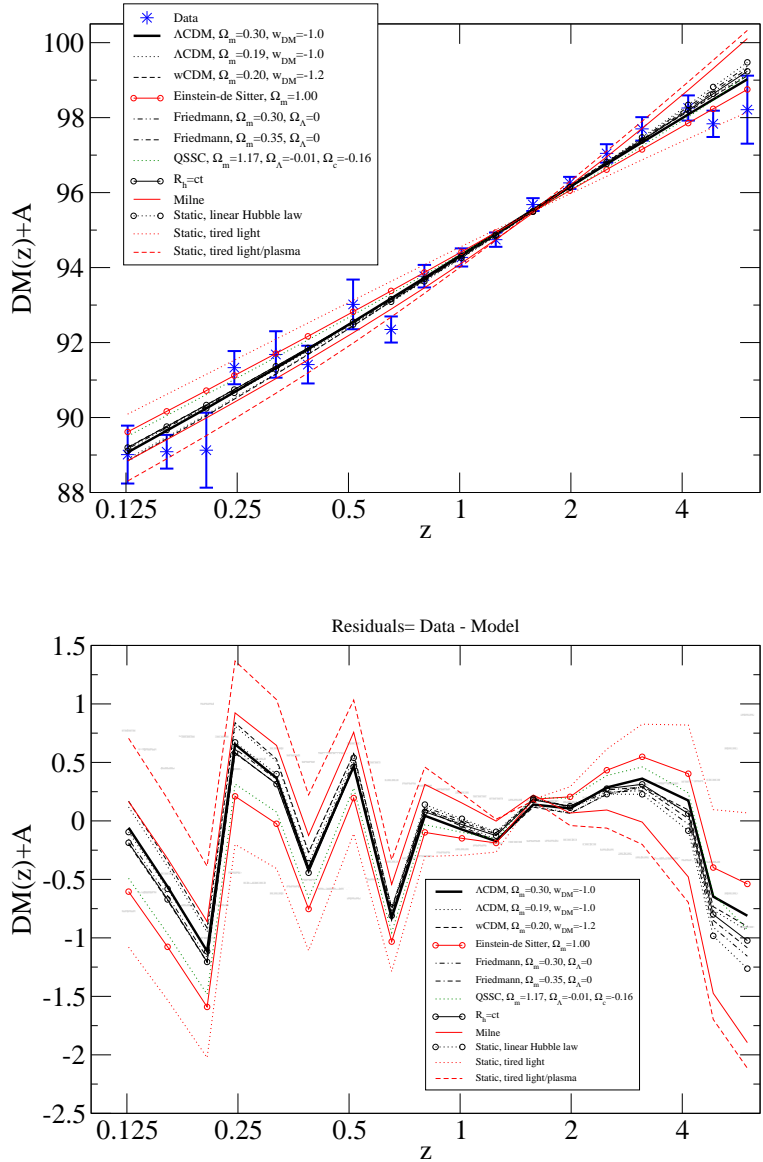

\centering
\vspace{.4cm}
\includegraphics[width=10cm]{fig1a.eps}\\
\vspace{1cm}
\includegraphics[width=10cm]{fig1b.eps}
\caption{Top: Log-linear plot of data and best fits for the distance modulus (+scaling constant $A$)
in various cosmological models (see Table
\ref{Tab:chi2}). Bottom: residuals $(Data-Model)$ corresponding to the above fits, for which the grey shaded region is the region within the error bar for
which Data and Model are coincident.
Black color indicates that the fit is good within 95\% C.L.; green color indicates
that the cosmological model is excluded at a C.L. between 95\% and 99\%; red color
indicates that the cosmological model is excluded at C.L. larger than 99\%.
Error bars only reflect the error of the average due to the dispersion of data; they do not contain the error in $\gamma $, which is fixed at $\gamma =0.60$.}
\vspace{0.5cm}
\label{Fig:DM}
\end{figure}

For this purpose, we adopt the data prepared by Ref. \cite{Ris15} and bin them in a weighted average
in intervals
of $\Delta \log _{10}z=0.1$, ensuring that there are $N\ge 4$ QSOs/bin. In total, we have
18 data points with averaged values of $-6.25[\log _{10}(F_{\rm X})-0.60\log _{10}
(F_{\rm UV})]$, and an error given by $\frac{r.m.s.}{\sqrt{N-1}}$ (see Fig.~\ref{Fig:DM}).
Using the dispersion of values to determine the error of the average is more accurate
than using the individual error bars for the fluxes from different sources that
are underestimated or unknown in some cases. As argued by Ref. \cite{Ris15}, the method chosen
to bin the data does not significantly affect the fits.

\eject

\begin{landscape}
\begin{table*}
\begin{center}
{\bf Table 1}. Results of the $\chi ^2$ test, best-fit free cosmological parameters, if any
($\nu $ is the number of free parameters), and associated probabilities.
The number of points is $N=18$. The model
number corresponds to the list in \S \ref{.cosmomodels}.
\vspace{5mm}
\begin{tabular}{lcclcc}
\label{Tab:chi2}
\qquad\qquad Model & $\nu $ & $\chi _{\rm dof,min}^2$ & \qquad Free cosmol. parameters & $A$ & Probability \\ \hline
(1)-$\Lambda $CDM $\Omega _m=0.3$, $w _\Lambda =-1$ & 1 & 1.26 & --- & 50.21 & 0.21 \\
(1)-$\Lambda $CDM $w _\Lambda =-1$, $\Omega _m$ free & 2 & 1.27 & $\Omega_m=0.19^{+0.20}_{-0.11}$ & 50.01 & 0.21 \\
(1)-wCDM both $\Omega _m$, $w_{de} $ free & 3 & 1.35 & $\Omega_m=0.20^{+0.24}_{-0.20}$, $w_{de}=-1.2^{+1.6}_{-\infty }$ & 49.94 & 0.16 \\
(2)-EdS & 1 & 1.98 & --- & 50.89 & $9.1\times 10^{-3}$  \\
(3)-Fr.neg.curv. $\Omega _m=0.3$ & 1 & 1.41 & --- & 50.39 & 0.12 \\
(3)-Fr.neg.curv. $\Omega _m$ free & 2 & 1.49 & $\Omega _m=0.35^{+0.28}_{-0.20}$ & 50.44 & 0.093 \\
(4)-QSSC $\Omega _m$, $\Omega_\Lambda \le 0$ free & 3 & 2.01 & $\Omega_m=1.17^{+0.42}_{-0.15}$, $\Omega_\Lambda =-0.01^{+0.01}_{-0.36}$ & 50.77 & 0.011 \\
(5)-$R_h=ct$ & 1 & 1.38 & --- & 50.40 & 0.14 \\
(6)-Milne & 1 & 2.23 & --- & 50.04 & $2.5\times 10^{-3}$ \\
(7)-St.lin.Hub. & 1 & 1.45 & --- & 50.31 & 0.10 \\
(8)-St.tir.l. & 1 & 3.82 & --- & 51.42 & $1.6\times 10^{-7}$ \\
(9)-St.pl.tir.l. & 1 & 3.23 & --- & 49.38 & $7.1\times 10^{-6}$ \\
 \hline
\end{tabular}
\end{center}
\end{table*}
\end{landscape}

\eject

\subsection{QSO HD on its own}

Table \ref{Tab:chi2} and Fig. \ref{Fig:DM} show the results of our fits to the
QSO data using the nine cosmologies introduced in \S~2. Some of these models have
fixed parameters; others have parameters that need to be optimized in producing
the best fits. Of the nine cases, five are excluded at the $>95\%$ C.L.. These are
Einstein-de Sitter, the Quasi-steady State, the Milne Universe, the Static Euclidean
model with simple tired light, and the Static Euclidean model with plasma tired light.
All of them, except Quasi-stady State, are excluded at the $>99\%$ C.L. The remaining 
four models (standard $\Lambda$CDM, $R_{\rm h}=ct$, the Friedmann open universe, and 
the Static model with a linear Hubble law) pass the test. If we optimize $\Omega _m$ 
in $\Lambda$CDM, the best fit is obtained with $\Omega _m=0.19^{+0.20}_{-0.11}$ 
($1\sigma$ error bars); and if we allow both $\Omega _m$ and $w_{de}$ to be free, 
then the best fit corresponds to the optimized parameter values 
$\Omega_m=0.20^{+0.24}_{-0.20}$, $w_{de}=-1.2^{+1.6}_{-\infty}$.
The error bars include the uncertainty in $\gamma $ (one of the constants
in Eq.~10).

\eject

\begin{landscape}

\begin{table*}
\begin{center}
{\bf Table 2.} Combined QSO HD + Alcock-Paczynski (AP) test. References for
the AP results are: Refs. \protect{\cite{Lop14b,Mel15c}}. For the Milne model, the
results of the AP test were not published, but we have calculated the
probability here using the same method and data as in Ref. \cite{Mel15c}. This
reference is denoted Ref. \cite{Mel15c}*.
\vspace{5mm}
\begin{tabular}{llccc}
\label{Tab:combined}
\qquad\qquad Model  & \qquad Free parameters & Probability AP & Source AP & Combined probability\\ \hline
(1)-$\Lambda $CDM $\Omega _m=0.3$, $w_\Lambda =-1$ & --- & 0.0066 & Ref. \cite{Mel15c} & 0.036 \\
(1)-$\Lambda $CDM $w _\Lambda =-1$, $\Omega _m$ free & $\Omega _m=0.20^{+0.06}_{-0.04}$ & 0.0095 & Ref. \cite{Mel15c} & 0.057  \\
(1)-wCDM $\Omega _m$, $w_{de} $ free & $\Omega_m=0.38^{+0.20}_{-0.19}$, $w_{de}=-0.28^{+0.52}_{-0.40}$ & --- & Ref. \cite{Mel15c} & 0.16 \\
(2)-EdS & --- & $2.7\times 10^{-4}$ & Ref. \cite{Lop14b} & $4.1\times 10^{-7}$  \\
(3)-Fr.neg.curv. $\Omega _m=0.3$ & --- & 0.0034 & Ref. \cite{Lop14b} & 0.0029 \\
(3)-Fr.neg.curv. $\Omega _m$ free & $\Omega _m=0.03^{+0.02}_{-0.03}$ & 0.024 & Ref. \cite{Lop14b} & 0.0016 \\
(4)-QSSC $\Omega _m$, $\Omega _\Lambda \le 0$ free & $\Omega _m=1.22^{+0.11}_{-0.09}$, $\Omega _\Lambda =0.0^{+0.0}_{-0.08}$ & 0.020 & Ref. \cite{Lop14b} & $1.4\times 10^{-3}$ \\
(5)-$R_h=ct$ & --- & 0.96 & Ref. \cite{Mel15c} & 0.21 \\
(6)-Milne & --- & $1.1\times 10^{-9}$ & Ref. \cite{Mel15c}* & $7.3\times 10^{-10}$ \\
(7)-St.lin.Hub. & --- & $1.4\times 10^{-5}$ & Ref. \cite{Lop14b} & $2.2\times 10^{-5}$ \\
(8)-St.tir.l. & --- & 0.96 & Ref. \cite{Mel15c} & $6.0\times 10^{-7}$ \\
(9)-St.pl.tir.l. & --- & 0.96 & Ref. \cite{Mel15c} & $2.3\times 10^{-5}$  \\
 \hline
\end{tabular}
\end{center}
\end{table*}

\end{landscape}
\eject

\subsection{QSO HD combined with the Alcock-Paczy\'nski test}

An application of the Alcock-Paczy\'nski (AP) test to cosmological models
can provide additional tight constraints,\cite{Lop14b,Mel15c} independently of the
present HD for QSOs.
Very importantly, the AP test is entirely independent of any galaxy
evolution and this is the reason for choosing it among other possible 
constraints using only the galaxy distribution. The reason for this 
independence is that the measured quantity---the ratio of observed 
angular size to radial/redshift size in the anisotropic two–point 
correlation function---depends only on the geometry of the Universe, 
provided the distribution of galaxies is spherical, which is always true
at any age of the Universe. The only sources of contamination in that 
measurement are the redshift distortions produced by the peculiar 
velocities of the galaxies, but there is a way to overcome them with 
the inclusion in the AP test of an observational signature with a 
sharp feature, such as the Baryonic Acoustic Oscillation (BAO) peak.\cite{Mel15c} 
We avoid combining these tests with the analysis of CMB data, first, 
because the latter has already been used by many other authors and we 
want to introduce new considerations that have not been adopted in 
previous papers; and second, because the CMB analysis is not free 
of interpretation and needs some modelling, for instance, with the
foreground contamination, that makes it dependent on considerations 
other than a pure cosmological geometry (see \S \ref{.intro}).
 In Table \ref{Tab:combined}, we list
the combined total probability for the QSO HD and AP analysis (based on
the most recent and most accurate baryon acoustic oscillation [BAO]
measurements in the data from Ref. \cite{Mel15c}), including
the best-fit parameters resulting from a combination of the Gaussian
distributions in both tests. This combined probability is calculated by
summing the $\chi ^2$'s and the number of degrees of freedom ($DF$) of
both tests and calculating the corresponding probability of this sum, i.e.,
\begin{eqnarray}
&\null&P_{1+2}=P(\chi ^2,DF)\nonumber\\
&\null&\chi ^2=\chi_1^2(P_1,DF _1)+\chi_2^2(P_2,DF _2)\\
&\null&DF=DF_1+DF_2\;.\nonumber
\end{eqnarray}

The combined test leaves only two models that are not excluded at $>95\%$
C.L.: $\Lambda$CDM with parameters different from the standard model
and $R_{\rm h}=ct$. The $R_{\rm h}=ct$ cosmology
has the advantage that it can produce a good fit without any free
parameters, whereas the parameter optimization ($\Omega_m=0.38$ and
$w_{de}=-0.28$) in $\Lambda$CDM produces some tension with the concordance
values $\Omega_m=0.3$ and $w_{de}\equiv w_\Lambda=-1$ (see also Ref. \cite{Mel15c}).
The variation of cosmological parameters in $\Lambda$CDM would thus
diminish the level of concordance with other cosmological data. On the
other hand, the $R_{\rm h}=ct$ Universe without any free parameters has
successfully passed all other cosmological tests applied to it thus far.
\cite{Mel13a,Mel13b,Mel14b,Mel15b,Mel15a,Wei13,Wei14b,Wei15}

\section{Discussion and Conclusion}

A HD diagram for QSOs can constrain cosmological models in ways that
many other tests, e.g., involving the use of Type Ia SNe, cannot,
since quasar spectra can be studied at redshifts which are not accesible for Type Ia SNe.
Unfortunately, the present sample of objects that may
be used for this study still results in distance moduli with a large
dispersion at high $z$, roughly of order one magnitude, that mitigates
the overall power of this test. Furthermore, this test assumes that the
relation between the X-ray and UV fluxes is independent of redshift,
which is reinforced by the good fits produced assuming a constant
$\gamma (z)=0.60$ for $z\lesssim 6$.\cite{Ris15} Nonetheless, a $\gamma (z)$ with small variations
in redshift is not excluded within the current error bars.

Another caveat in this analysis is that there may be some unaccounted
for systematics associated with the determination of the K-corrections
used to calculate the X-ray flux in the rest frame (Ref. \cite{Ris15}, \S A.5).
Nonetheless, provided that these systematics and any possible evolution
in the relation between the X-ray and UV fluxes are negligible, one
can use present-day data to check whether a cosmological model predicts
a distance modulus in line with the QSO observations over a wide range in
redshift.

Ref. \cite{Ris15} was able to constrain the cosmological parameters in the
standard model with the available data. With these same data, we
demonstrated in this {\it letter} that five of nine different
cosmological models can be excluded at $>95\%$ C.L. These models
are: the Quasi-steady State model, Einstein-de Sitter, the Milne Universe,
the Static Euclidean model with simple tired light and the Static
Euclidean model with plasma tired light. These last four are
excluded very strongly, at $>99\%$ C.L.

The remaining four models [standard $\Lambda$CDM, $R_{\rm h}=ct$ (with
zero active mass), the Friedmann open model and the Static model with
a linear Hubble law] all pass the QSO HD test, but only the first two
also pass the Alcock-Paczy\'nski test using the latest, high-precision
BAO data. Future surveys will increase the QSO sample suitable for
this study and permit a determination of the X-ray K-correction
directly from their spectra, with the possibility of further increasing
the precision of the QSO HD and eliminating additional models from
the list in \S~2. At the same time, this diagnostic will continue
to refine the optimization of parameters in the standard model.


\section*{Acknowledgments}
Thanks are given to the anonymous referee for helpful comments.
FM is grateful to Amherst College for partial support through
the John Woodruff Simpson Chair. 
EL and GR have been supported by the grant PRIN-INAF 2012.

\end{document}